\documentclass[runningheads]{llncs}
\usepackage[utf8]{inputenc}
\usepackage{graphicx}
\usepackage{amsmath}
\usepackage{lipsum}
\usepackage{booktabs}
\usepackage{threeparttable}
\usepackage{makecell}
\usepackage{bbding}
\usepackage{amssymb}
\usepackage{adjustbox}
\usepackage{cite}
\usepackage{float}
\newcommand{\ucite}[1]{\cite{#1}}

\begin{document}
\title{DiffULD: Diffusive Universal Lesion Detection}
%
%
\author{Peiang Zhao\inst{1} \and Han Li\inst{1} \and
Ruiyang Jin\inst{1} \and S. Kevin Zhou\inst{1,2}}
%
\authorrunning{P. Zhao et al.}
%
\institute{School of Biomedical Engineering \& Suzhou Institute for Advanced Research, Center for Medical Imaging, Robotics, and Analytic Computing \& LEarning (MIRACLE), University of Science and Technology of China, Suzhou 215123, China\\
\and
Key Lab of Intelligent Information Processing of Chinese Academy of Sciences (CAS), Institute of Computing Technology, CAS, Beijing, 100190, China}
%
\maketitle              
\begin{abstract}

Universal Lesion Detection (ULD) in computed tomography (CT) plays an essential role in computer-aided diagnosis. Promising ULD results have been reported by anchor-based detection designs, but they have inherent drawbacks due to the use of anchors: \textbf{i) Insufficient training target }and \textbf{ii) Difficulties in anchor design.} Diffusion probability models (DPM) have demonstrated outstanding capabilities in many vision tasks. Many DPM-based  approaches achieve great success in natural image object detection without using anchors. But they are still ineffective for ULD due to the insufficient training targets.   
In this paper, we propose a novel ULD method, DiffULD, which utilizes DPM for lesion detection. To tackle the negative effect triggered by insufficient targets, we introduce a novel center-aligned bounding box padding strategy that provides additional high-quality training targets yet avoids significant performance deterioration. DiffULD is inherently advanced in locating lesions with diverse sizes and
shapes since it can predict with arbitrary boxes. 
Experiments on the benchmark dataset DeepLesion\cite{yan2018deeplesion} show the superiority of DiffULD when compared to state-of-the-art ULD approaches.
\end{abstract}
\section{Introduction}
Universal Lesion Detection (ULD) in computed tomography (CT)\cite{zlocha2019improving,tao2019improving,zhang20193d,zhang20203d,tang2019uldor,yan183dce,li2019mvp,yan2019mulan,yang2020alignshift,cai2020deep,li2020bm,zhang2020revisiting,yan2020learning,tang2021weakly,yang2021asymmetric,li2021conditional,lyu2021segmentation} plays an important role in computer-aided diagnosis (CAD)\ucite{zhou2019handbook,zhou2021review}.  The design of detection-only instead of identifying the lesion types in ULD\ucite{yu2020deep,ren2021retina,baumgartner2021nndetection,shahroudnejad2021tun,luo2021oxnet,chen2021ellipsenet,lin2021automated,zhao2021positive} prominently decreases the difficulty of this task for a specific organ (e.g., lung, liver), but it is still challenging for lesions vary in shapes and sizes among whole human body.

Previous arts in ULD are mainly motivated by the anchor-based detection framework, \textit{e.g.}, Faster-RCNN \cite{renNIPS15fasterrcnn}.
These studies focus on adapting the detection backbone to universally locate lesions in CT scans. 
For instance, Li \textit{et al.}~\cite{li2019mvp} propose the so-called MVP-Net, a multi-view FPN with a position-aware attention mechanism to assist ULD training.
Yang \textit{et al.}~\cite{yang2019ACS,yang2020alignshift,yang2021asymmetric} propose a series of 3D feature fusion operators to incorporate context information from several adjacent CT slices for better performance. 
Li \textit{et al.}~\cite{li2022satr} introduce a plug-and-play transformer block to form hybrid backbones which can better model long-distance feature dependency.

\begin{figure}[htb]
\includegraphics[width=0.95\textwidth]{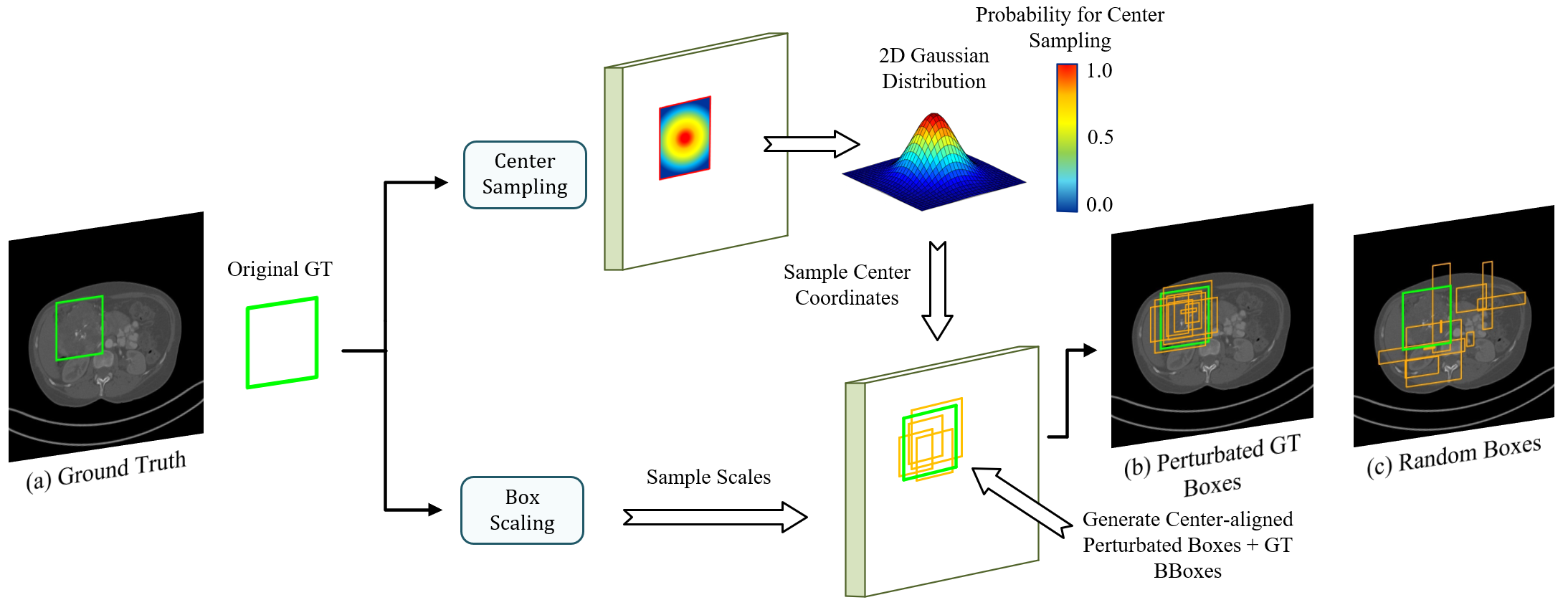}
\caption{Overview of the proposed center-aligned BBox padding strategy.}\label{fig1}
\end{figure}

While achieving success, these anchor-based methods have inherent drawbacks:
(i) \textbf{Insufficient training target.}

In stage-1, anchor-based methods identify the positive (lesion) anchors and label them as the region of interest (RoI)  based on the IoU between anchors and ground-truth (GT) bounding boxes (BBoxes). An anchor is considered positive if its IoU with any GT BBox is greater than the IoU threshold and negative otherwise\cite{renNIPS15fasterrcnn}.
The positive anchors are sufficient in natural images as they usually have many targets per image \cite{li2020bm}.
However, the number of lesions per CT scan is limited, most CT slices only contain one or two lesions (\textit{i.e.}, detection targets in ULD) per CT slice \cite{li2021conditional}. Still applying the IoU-based anchor matching mechanism with such limited targets can lead to severe data imbalance and further hinders network convergence.
Simply lowering the positive IoU threshold in the anchor-selecting mechanism can alleviate the shortage of positive anchors to some degree, but it leads to a higher false positive (FP) rate by labeling more low-IoU anchors as positive.
(ii) \textbf{Difficulties in anchor design.} 
In anchor-based methods, the size, ratio and number of anchors are pre-defined hyper-parameters that significantly influence the detection performance\cite{sheoranBMVCdkma}. Thus a proper design of anchor hyper-parameters is of great importance. However, tuning anchor hyper-parameters is a challenging task in ULD because of the variety of lesions (target) with diverse diameters (from 0.21 to 342.5mm). Even with a careful design, the fixed rectangle anchor boxes can be a kind of obstruction in capturing heterogeneous lesions, which have irregular shapes and vary in size.

To get rid of the drawbacks caused by the anchor mechanism, researchers resort to anchor-free detection, \textit{e.g.}, FCOS\cite{tian2019fcos} and DETR\cite{carion2020end}. But these methods experience difficulties in achieving state-of-the-art results in ULD, as they lack the initialization of position prior provided by anchors. 

Recently, the diffusion probabilistic model (DPM)~\cite{jo2020dpm,sahar2022imageen,chen2022generalist,holmquist2022diffpose,kim2022diffusemorph} has demonstrated its outstanding capabilities in various vision tasks. Chen \textit{et. al.} follow the key idea of DPM and propose a noise-to-box pipeline, DiffusionDet\cite{chen2022diffusiondet}, for natural image object detection. They achieved success on natural images with sufficient training targets, but still experience difficulties in dealing with tasks with insufficient training targets like ULD.
This is because the DPM's denoising is a dense distribution-to-distribution forecasting procedure that heavily relied on a large number of high-quality training targets to learn targets' distribution accurately.

To address this issue, we hereby introduce a novel center-aligned BBox padding strategy in DPM detection to form a diffusion-based detector for Universal Lesion Detection, termed DiffULD.

As shown in Fig. 1, DiffULD also formulates lesion detection as a denoising diffusion process from noisy boxes to prediction boxes similar to \cite{chen2022diffusiondet}, but we further introduce the center-aligned BBox padding before DiffULD's forward diffusion process to generate perturbated GT. 
Specifically, we add random perturbations to both scales and center coordinates of the original GT BBox, resulting in \textit{perturbated boxes} whose centers remain aligned with the corresponding original GT BBox. 
Next, original GT boxes paired with these perturbated boxes are called \textit{perturbated GT boxes} for simplicity. 
Finally, we feed the perturbated GT boxes to the model as the training objective during training.
Compared with other training target padding methods (\textit{e.g.}, padding with random boxes), our strategy can provide additional targets of higher quality, \textit{i.e.}, center aligned with the original GT BBox. This approach effectively expands the insufficient training targets on CT scans, enhancing  DPM's detection performance and avoiding deterioration triggered by adding random targets. 


The following DPM training procedure contains two diffusion processes. i) In the forward training process, DiffULD corrupts the perturbated GT with Gaussian noise gradually to generate noisy boxes step by step. Then the model is trained to remove the noise and reconstruct the original perturbated GT  boxes. ii) In the reverse inference process, the trained DiffULD can refine a set of randomly generated boxes iteratively to obtain the final detect predictions.

Our method gets rid of the drawbacks of pre-defined anchors and the deterioration of training DPM with insufficient GT targets. Besides, DiffULD is inherently advanced in locating targets with diverse sizes since it can predict with arbitrary boxes, which is an advantageous feature for detecting lesions of irregular shapes and various sizes. To validate the effectiveness of our method, we conduct experiments against seven state-of-the-art ULD methods on the public dataset DeepLesion\cite{yan2018deeplesion}. The results demonstrate that our method achieves competitive performance compared to state-of-the-art ULD approaches.

\section{Method}
In this section, we first formulate our overall detection process for DiffULD and then specify the training manner, inference process and backbone design.
\subsection{Diffusion-based detector for lesion detection}
Universal Lesion Detection can be formulated as locating lesions in input CT scan $I_{ct}$ with a set of boxes predictions ${z^{'}_{0}}$. For a particular box ${z^{'}}$, it can be denoted as ${z^{'}}=[{x}_{1}, {y}_{1},{x}_{2}, {y}_{2}]$, where ${x}_{1}, {y}_{1}$ and ${x}_{2}, {y}_{2}$ are the coordinates of the top-left and bottom-right corners, respectively.

We design our model based on a diffusion model mentioned in \cite{chen2022diffusiondet}. As shown in Fig. 2, our method consists of two stages, a forward diffusion (or training) process and a reverse refinement (or inference) process. 
In the forward process, We denote GT bounding boxes as ${z}_{0}$ and generate corrupted training samples ${z}_{1},{z}_{2},...,{z}_{T}$
for latter DiffULD training by adding Gaussian noise iteratively, which can be defined as:
\begin{equation}
\left({z}_{t} \mid {z}_{0}\right)=\mathcal{N}\left({z}_{t} \mid \sqrt{\bar{\alpha}_{t}} {z}_{0},\left(1-\bar{\alpha}_{t}\right){I}\right)
\end{equation}
where $\bar{\alpha}_{t}$ represents the noise variance schedule and $t \in \{0, 1, ..., T \}$. Subsequently, a neural network ${f}_{\theta}(z_{t}, t, I_{ct})$ conditioned on the corresponding CT scan $I_{ct}$ is trained to predict $z_{0}$ from a noisy box ${z}_{T}$ by reversing the noising process step by step. During inference, for an input CT scan $I_{ct}$ with a set of random boxes, the model is able to refine the random boxes to get a lesion detection prediction box ${z^{'}_{0}}$, iteratively.
\begin{figure}[tb]
\includegraphics[width=\textwidth]{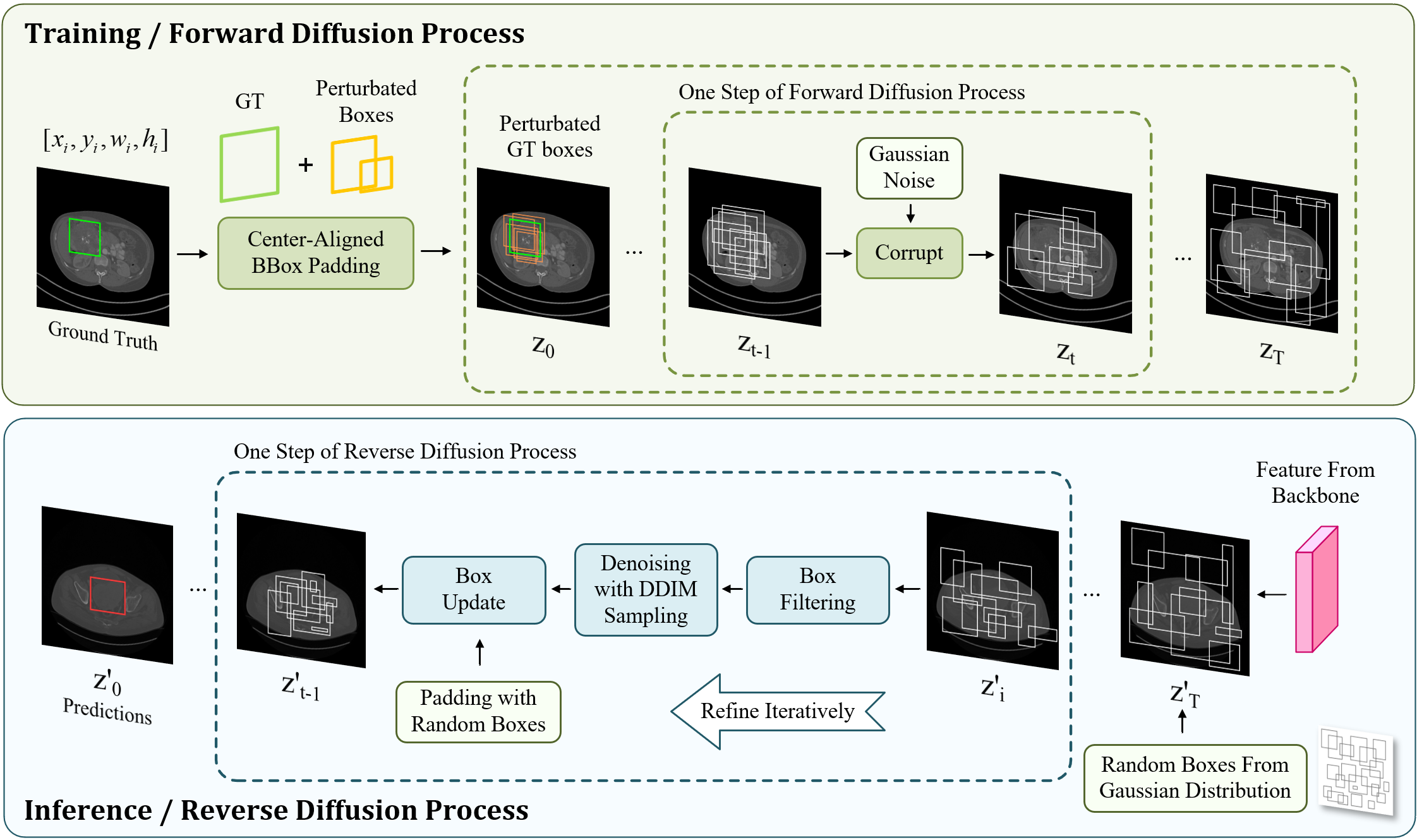}
\caption{Overview of DiffULD. The backbone extracts feature representation from an input CT scan. Then, the model takes the feature map and a set of noisy boxes as input and makes predictions.}\label{fig2}
\end{figure}
\subsection{Training}
In this section, we specify the training process with our novelty introduced `Center-aligned BBox padding'.


\textbf{Center-aligned BBox padding.} 
As shown in Fig. 1, we utilize Center-aligned BBox padding to generate \textit{perturbated boxes}. Then, the perturbated boxes are paired with original GT BBoxes, forming \textit{perturbated GT boxes} which are used to generate corrupted training samples ${z}_{1},{z}_{2},...,{z}_{T}$
for the latter DiffULD training by adding Gaussian noise iteratively. 

For clarity, we denote $[x_{c}^{i},y_{c}^{i}]$ as center coordinates of original GT BBox, where $z^{i}$, $[w^{i},h^{i}]$ are the width and height of $z^{i}$.

We consider the generation in two parts: box scaling and center sampling. 
(i) \textit{Box scaling:} We set a hyper-parameter $\lambda_{scale}\in(0,1)$ for scaling. For $z^{i}$, The width and height of the corresponding perturbated boxes are randomly sampled in uniform distributions on $U_{w}\sim[(1-\lambda_{scale})w^{i}, (1+\lambda_{scale})w^{i}]$ and $U_{h}\sim[(1-\lambda_{scale})h^{i}, (1+\lambda_{scale})h^{i}]$.
(ii) \textit{Center sampling:} We sample the center coordinates $[{x}_{c}^{new}, {y}_{c}^{new}]$ of perturbated boxes from a 2D Gaussian distribution $\mathcal{N}$ whose probability density function can be denoted as:
\begin{equation}
f_{}(x,y)=
\exp \left(-\frac{\left(x-{x}_{c}^{i}\right)^{2}+\left(y-{y}_{c}^{i}\right)^{2}}{2 \sigma^{2}}\right), ~~\quad (x, y) \in z_{i}
\end{equation}
where $\sigma$ is a size-adaptive parameter, which can be calculated according to the $z^{i}$'s width and height:
\begin{equation}
    \sigma=\frac{1}{6}(w^{i}+h^{i}).
\end{equation}
Besides, for each input CT scan $I_{ct}$, we collect all GT BBoxes in ${z}$ and add random perturbations to  them and generate multiple \textit{perturbated boxes} for each of them. Thus the number of perturbated boxes in an image varies with its number of GT BBoxes. For better training, we fix the number of perturbated boxes as $N$ for all training images.

As shown in Fig 1., the perturbated boxes cluster together and their centers are still aligned with the corresponding original GT BBox. Subsequently, perturbated GT boxes $z_{0}$ are sent for corruption as the training objective.

\textbf{Box corruption.} As shown in Fig. 1, we corrupt the parameters of $z_{0}$ with Gaussian noises. The noise scale is controlled by $\bar{\alpha}_{t}$ (in Eq. 1), which adopts the decreasing cosine scheduler in the different time step $t$.

\textbf{Loss function.} As the model generates the same number of ($N$) predictions for the input image, termed as a prediction set, the loss function should be set-wised \cite{carion2020end}. Specifically, each GT is matched with the prediction by the least matching cost, and the overall training loss\cite{carion2020end} can be represented as:
\begin{equation}
    \mathcal{L}=\lambda_{L 1 box} \cdot \mathcal{L}_{L 1box}+\lambda_{giou} \cdot \mathcal{L}_{giou }
\end{equation}
where $\mathcal{L}_{L 1box}$ and  $\mathcal{L}_{giou }$ are the pairwise box loss. We adopt $\lambda_{L 1 box}=2.0$ and $\lambda_{giou }=5.0$.
\subsection{Inference}
At the inference stage, with a set of random boxes sampled from Gaussian distribution, the model does refinement step by step to obtain the final predictions $z^{'}_{0}$. For better performance, two key components are used: 

\textbf{Box filtering.} In each refinement step, the model receives a set of box proposals from the last step. As the prediction starts from arbitrary boxes and the lack of GT (lesion), most of these proposals are very far from lesions. Keeping refining them in the following steps will hinder network training. Toward efficient detection, we send the proposals to the detection head and remove the boxes whose confidential scores are lower than a particular threshold $\lambda_{conf}$.
The remaining high-quality proposals are sent for followed DDIM sampling.



\textbf{Box update with DDIM sampling.} DDIM\cite{song2020denoising} is utilized to further refine the received box proposals by denoising. Next, these refined boxes are sent to the next step and start a new round of refinement. After multiple steps, final predictions are obtained.

However, we observe that if we just filter out boxes with low scores during iterative refinement, the model runs out of usable box proposals rapidly, which also leads to a deterioration in performance. Therefore, after the box updating, we add new boxes sampled from a Gaussian Distribution to the set of remaining boxes with. The number of box proposals per image is padded to the fixed number $N$ before they are sent to the next refinement step.

\subsection{Backbone Design}


\textbf{Multi-window input.} 
Most prior arts in ULD use a single and fixed window (\textit{e.g.}, a wide window of [1024, 4096]) to render the input CT scan, which suppresses organ-specific information and makes it hard for the network to focus on the various lesions.
Therefore, taking cues from \cite{li2019mvp}, we introduce 3 organ-specific HU windows to highlight multiple organs of interest. Their window widths and window levels are: ${W}_{1} = [1200, -600]$ for chest organs, ${W}_{2} = [400, 50]$ for soft tissues and ${W}_{3} = [200, 30]$ for abdomenal organs. Multi-window features are extracted with a ConvNeXt-T shared network.

\textbf{3D context feature fusion.}
We modify the original A3D\cite{yang2021asymmetric} DenseNet backbone for context fusion. We remove the first Conv3D Block and use the truncated network as our 3D context fusion module, which fuses the multi-view features from the last module. Multi-window features are fused with this module and sent to the detector subsequently for lesion detection.

\section{Experiments}
\subsection{Settings}
Our experiments are conducted on the standard ULD dataset DeepLesion\cite{yan2018deeplesion}. The dataset contains 32,735 lesions on 32,120 axial slices from 10,594 CT studies of 4,427 unique patients. We use the official data split of DeepLesion which consists of 70\%, 15\%, 15\% for training, validation, and test, respectively. Besides, we also evaluate the performance of 3 methods based on a revised test set from \cite{}.

\textbf{Training details.} DiffULD is trained on CT scans of size 512 × 512 with a batch size of 4 on 4 NVIDIA RTX Titan GPUs with 24GB memory.  For hyper-parameters, the threshold $N$ for box padding is set to $300$. $\lambda_{scale}$ for box scaling is set to 0.4. $\lambda_{conf}$ for box filtering is set to 0.5. We use the Adam optimizer with an initial learning rate of $2e-4$ and the weight decay as $1e-4$. The default training schedule is 120K iterations, with the learning rate divided by 10 at 60K and 100K iterations. Data augmentation strategies contain random horizontal flipping, rotation, and random brightness adjustment.

\textbf{Evaluation metrics.}
The lesion detection is classified as true positive (TP) when the IoU between the predicted and the GT bounding box is larger than 0.5. Average sensitivities computed at 0.5, 1, 2, and 4 false-positives (FP) per image are reported as the evaluation metrics on the test set for a fair comparison. 
\begin{table}[tb]
\centering
\caption{Sensitivity (\%) at various FPPI on the standard test set of DeepLesion. DKA-ULD\cite{sheoran2022efficient} and SATr\cite{li2022satr} are up-to-date SOTA ULD methods under the settings of 3 slices and 7 slices, respectively. }\label{tab1}
\begin{adjustbox}{center}
\scalebox{0.75}{
\footnotesize{
\renewcommand{\arraystretch}{1.1}
\setlength{\tabcolsep}{1.5mm}{}

\begin{tabular}{p{75pt}cccccc}
\toprule
\textbf{Methods} & \textbf{Slices} & \textbf{@0.5} & \textbf{@1} & \textbf{@2} & \textbf{@4} & \textbf{Avg.[0.5,1,2,4]} \\ \midrule
3DCE\cite{yan183dce}             & 27                             & 62.48        & 73.37      & 80.70     & 85.65      & 75.55                    \\
RetinaNet\cite{zlocha2019improving}        & 3                            & 72.18         & 80.07       & 86.40        & 90.77       & 82.36                        \\
MVP-Net\cite{li2019mvp}          & 9                               & 73.83         & 81.82       & 87.60        & 91.30        & 83.64                        \\
MULAN\cite{yan2019mulan}            & 27                              & 76.10          & 82.50        & 87.50        & 90.90        & 84.33                        \\
AlignShift\cite{yang2020alignshift}       & 3                               & 73.00            & 81.17       & 87.05       & 91.78       & 83.25                        \\
A3D\cite{yang2021asymmetric}              & 3                               & 74.10          & 81.81       & 87.87       & 92.13       & 83.98                        \\
\midrule
DKA-ULD\cite{sheoran2022efficient}          & 3                             & 77.38         & 84.06       & 89.28       & 92.04     & 85.79                        \\ 
DiffULD (Ours)        & 3                           & \textbf{77.84} (0.46$\uparrow$) & \textbf{84.57} (0.51$\uparrow$) & \textbf{89.41} (0.13$\uparrow$)& \textbf{92.31} (0.40$\uparrow$)          & \textbf{86.03} (0.58$\uparrow$)                       \\
\midrule

SATr\cite{li2022satr}        & 7                             & \textbf{81.02}         & 86.64       & 90.69       & \textbf{93.30}    & 87.91                        \\
DiffULD (Ours)        & 7                            &80.43 & \textbf{87.16} (0.52$\uparrow$)  & \textbf{91.20} (0.51$\uparrow$)  & 93.21         & \textbf{88.00} (0.08$\uparrow$)         \\\midrule
FCOS\cite{tian2019fcos}           & 3                               & 56.12          & 67.31       & 73.75       & 81.44      & 69.66                        \\
CenterNet++\cite{duan2022centernet}            & 3                               & 67.34         & 75.95       & 82.73       & 87.72       & 78.43                        \\
DN-DETR\cite{li2022dndetr}           & 3                               & 69.27          & 77.90      & 83.97       & 88.59       & 81.02    \\                    
\bottomrule
\end{tabular}
}}
\end{adjustbox}
\end{table}

\begin{table}[tb]
\centering
\caption{Sensitivity (\%) at various FPPI on the augmented test set introduced by Lesion-Harvester\cite{cai2020lesion}.}\label{tab2}
\begin{adjustbox}{center}
\scalebox{0.75}{
\footnotesize{

\renewcommand{\arraystretch}{1.1}
\setlength{\tabcolsep}{1.5mm}{}
\begin{tabular}{p{75pt}cccccc}
\toprule
\textbf{Methods} & \textbf{Slices} & \textbf{@0.5} & \textbf{@1} & \textbf{@2} & \textbf{@4} & \textbf{Avg.[0.5,1,2,4]} \\ \midrule
A3D\cite{yang2021asymmetric}             & 7                             & 88.73        & 91.23      & 93.89    & 95.91      & 92.44                    \\
SATr\cite{li2022satr}        & 7                            & 91.04         & 93.75       & 95.58       & 96.73     & 94.28                        \\
DiffULD (Ours)          & 7                             & \textbf{91.65} (0.61$\uparrow$)         & \textbf{94.33} (0.58$\uparrow$)      & \textbf{95.97} (0.39$\uparrow$)       & \textbf{96.86} (0.13$\uparrow$)       & \textbf{94.70} (0.42$\uparrow$)                      \\
\bottomrule
\end{tabular}

}
}
\end{adjustbox}
\end{table}
\subsection{Lesion detection performance}
We evaluate the effectiveness of DiffULD against anchor-based ULD approaches such as 3DCE\cite{yan183dce}, MVP-Net\cite{li2019mvp}, A3D\cite{yang2021asymmetric} and SATr\cite{li2022satr} on DeepLesion. Several anchor-free natural image detection methods such as FCOS\cite{tian2019fcos} and DN-DETR\cite{li2022dndetr} are also introduced for comparison.
In addition, we conduct an extensive experiment to explore DiffULD's potential on an augmented test set of completely annotated DeepLesion volumes, introduced by Lesion-Harvester\cite{cai2020lesion}.

Table 1. demonstrates that our proposed DiffULD achieves favorable performances when compared to recent state-of-the-art anchor-based ULD approaches such as SATr on both 3 slices and 7 slices. 
It outperforms prior well-established methods such as A3D and MULAN by a non-trivial margin. This validates that with our padding strategy, the concise DPM can be utilized in general medical object detection tasks such as ULD and attain impressive performance.

\subsection{Ablation study}
\begin{table}[tb]
\centering
\caption{Ablation study of padding strategies at various FPs per image (FPPI).}\label{tab3}
\begin{adjustbox}{center}
\scalebox{1}{
\scriptsize{
\renewcommand{\arraystretch}{1.1}

\setlength{\tabcolsep}{1.3mm}{}
\begin{tabular}{cccccccc}
\toprule
Baseline & \makecell{Duplicate} & \makecell{Gaussian} & \makecell{Uniform} & \makecell{Center-Aligned} & FPPI = 0.5 & FPPI = 1 \\\midrule
\checkmark        &                  &                &                                                            &      & 76.71    & 83.49 \\
    \checkmark   &      \checkmark             &                &                                                                  & & 77.01    & 83.61  \\
        \checkmark      &              &       \checkmark              &                                                              &    & 77.68    & 83.98  \\
      \checkmark        &              &                 &      \checkmark                                                              &  & 77.22      & 83.75 \\
         \checkmark                 &              &                 &                         &                \checkmark                           & \textbf{77.84} (1.13$\uparrow$)    & \textbf{84.57 }(1.08$\uparrow$)   \\
\bottomrule
\end{tabular}

  }
}
\end{adjustbox}
\end{table}
We provide an ablation study about our proposed approach: center-aligned BBox padding. As shown in Table 3., we compared it with various other padding strategies, including: (i) duplicating original GT boxes; (ii) padding random boxes sampled from a uniform distribution; (iii) padding random boxes sampled from a Gaussian distribution; (iv) padding with center-aligned strategy.

Our baseline method is training the diffusion model directly with no box padding, using our proposed backbone in 2.4. The performance is increased by 0.30\% by simply duplicating the original GT boxes. Padding random boxes following uniform and Gaussian distributions brings 0.51\% and 0.91\% improvement respectively. Our center-aligned padding strategy accounts for 1.13\% improvement from the baseline. 
We attribute this performance boost to center-aligned padding's ability to provide high-quality additional training targets. It effectively expands the insufficient training targets on CT scans and enhances DPM's detection performance while avoiding deterioration triggered by adding random targets. This property is favorable for utilizing DPMs on a limited amount of GT like ULD.

\section{Conclusion}
In this paper, we propose a novel ULD method termed DiffULD by introducing the diffusion probability model (DPM) to Universal Lesion Detection. 
We present a novel center-aligned BBox padding strategy to tackle the performance deterioration caused by directly utilizing DPM on CT scans with sparse lesion bounding boxes. 
Compared with other training target padding methods (e.g., padding with random boxes), our strategy can provide additional training targets of higher quality and boost detection performance while avoiding significant deterioration.
DiffULD is
inherently advanced in locating lesions with diverse sizes and shapes since it can predict with
arbitrary boxes, making it a promising method for ULD.
Experiments on both standard and augmented DeepLesion datasets show that our proposed method can achieve competitive performance compared to state-of-the-art ULD approaches.
\clearpage
\bibliographystyle{unsrt}
\bibliography{ref}
\end{document}